\documentclass[conference]{IEEEtran}
\IEEEoverridecommandlockouts
\usepackage{amsmath,amssymb,amsfonts}
\usepackage{algorithmic}
\usepackage{graphicx}
\usepackage{textcomp}
\usepackage{xcolor}
\usepackage{textalpha}

\usepackage[style=numeric,sorting=none]{biblatex}
\def\BibTeX{{\rm B\kern-.05em{\sc i\kern-.025em b}\kern-.08em
    T\kern-.1667em\lower.7ex\hbox{E}\kern-.125emX}}

\begin{document}

\title{\vspace{-1.5cm} 
    \begin{center}
        \normalsize\normalfont\itshape
        This is the peer-reviewed version of the following article: Spatial model of neurogliovascular unit: arachidonic acid can control the stability, which has been published in final form at https://doi.org/10.1109/DCNA63495.2024.10718464. © 2024 IEEE. Personal use of this material is permitted. Permission from IEEE must be obtained for all other uses.
    \end{center}
    \vspace{0.2cm}
    \rule{\textwidth}{0.5pt} 
    \vspace{0.8cm}
    \\ 
    Spatial model of neurogliovascular unit: arachidonic acid can control the stability\\
\thanks{This study was supported by the Russian Science Foundation, grant 22-74-00146}
}

\author{\IEEEauthorblockN{Darya\,V.\,Verveyko}
\IEEEauthorblockA{\textit{Department of Theoretical Physics} \\
\textit{Kursk State University}\\
Kursk, Russia \\
allegroform@mail.ru}
\and
\IEEEauthorblockN{Andrey\,Yu.\,Verisokin}
\IEEEauthorblockA{\textit{Department of Theoretical Physics} \\
\textit{Kursk State University}\\
Kursk, Russia \\
ffalconn@mail.ru}
\and
\IEEEauthorblockN{Pavel\, O.\,Lukin}
\IEEEauthorblockA{\textit{Department of Theoretical Physics} \\
\textit{Kursk State University}\\
Kursk, Russia \\
luckinpavel97@gmail.com}
}

\maketitle

\begin{abstract}
Arachidonic acid metabolites can have both vasoconstrictor and vasodilator effects, thereby influencing neuronal activity. In this work, to analyze possible patterns of activity of nervous tissue in response to changes in the metabolic profile, a mathematical model of a spatially distributed system of connected neurogliovascular units was developed. We simulate the influence of arachidonic acid metabolites on the dynamics of the neurogliovascular unit during cortical spreading depression in order to determine the ways in which astrocytic activity influences the spatiotemporal characteristics of the propagating cortical depression wave. It is assumed that in the future the results of the theoretical study can be used to find ways to remove nervous tissue from pathological conditions that occur in epilepsy, migraines and a number of neurodegenerative diseases, including those associated with the occurrence of cortical depression waves.

\end{abstract}


\section{Introduction}
Recently experimental data confirm the presence of the vasomodulatory nature of arachidonic acid (AA) metabolites. It has been shown that modulation of the activity of the vasomotor muscles has an indirect effect on neural activity, determining the concentration of available oxygen and metabolic energy reserves necessary to maintain ion gradients and, in particular, the operation of sodium-potassium pumps, and therefore affects the functioning of the brain as a whole. Signal transmission between neurons and astrocytes through neurogliotransmitter interaction, in particular through the release of glutamate, leads to the release of AA metabolites from astrocytes, which carry out complex bipolar regulation of vascular tone (constriction/dilation)~\cite{Koehler2006}, while increasing the concentration of glial calcium and the subsequent formation of AA derivatives has an indirect effect on the activity of neighboring neurons~\cite{Attwell:2010}. 

It has been shown that the synthesis of AA derivatives depends on the local oxygen concentration in nerve tissues, and different AA metabolites have different effects on the dynamics of blood flow. Prostaglandin E$_2$ (PGE$_2$) and epoxyeicosatrienoic acids (EETs) lead to dilation of blood vessels, and 20-hydroxyeicosatetraenoic acid (20-HETE) causes vasoconstriction. Free AA may be involved in the development of Alzheimer's disease through various mechanisms. Free AA is also involved in synaptic functions as a retrograde messenger and regulator of neurotransmitter exocytosis.


Currently, there are a large number of models that describe neural activity, calcium waves in astrocytes, interactions between neurons and astrocytes, and neurovascular connections with varying degrees of detail (see, for example, the review~\cite{Manninen:2018}). However, the vast majority of recent models do not address the role of AA metabolites synthesized in astrocytes, limiting themselves to modeling calcium waves in astrocytes or neuronal activity. At the same time, experimental data allow us to confidently assert that AA and NO dynamics are responsible for controlling blood flow and neural activity in critical situations accompanied by hypoxia or hyperoxia.

In this work, a spatial model of the neurogliovascular unit (NGVU) was constructed taking into account the vasomodulatory effect of AA metabolites. We include in our consideration the level of available oxygen, which affects the synthesis of AA metabolites, which subsequently have a direct effect on the radius of blood vessels. 
Numerical solutions of local models were carried out by the Adams\,--\,Bashforth method; in the case of a spatial solution of the model, it was performed in the AGEOM~\cite{AGEOM_CUDA_SOFTWARE} environment using the 4th order Runge\,--\,Kutta method.


\section{Modeling}

The spatial model proposed in the work takes into account a wide range of processes occurring within the connected NGVUs. Thus, the model includes: (i) IP$_3$-dependent calcium dynamics; (ii) synaptic activity; (iii) functioning of sodium-potassium channels; (iv) the level of ATP, which determines metabolic energy reserves; (v) the dependence of the rate of cerebral oxygen metabolism on the state of NGVU. Below one can find the main equations of the model.

As a basis, we take the local version of the model designed earlier for modeling calcium activity in an astrocyte~\cite{Verisokin:2021}.

In this case, the dependence of the intensity of calcium events on glutamate is determined by the value of $J_{Glu}$, and the value of glutamate depends on the current level of synaptic activity, determined by ${V_{{K^ + }}}$. To simplify the study, we use a minimalistic model of neural activity~\cite{Chizhov:2018} through a description of the functioning of sodium-potassium pumps:



\begin{eqnarray}
\frac{{d[{K^ + }]}}{{dt}} = \frac{{{{[{K^ + }]}_{bath}} - [{K^ + }]}}{{{\tau _{{K^ + }}}}} - 2\lambda {I_{p}} + {\delta _{{K^ + }}}{V_{{K^ + }}} + {J_{{K_{ext}}}},\\
\frac{{d[N{a^ + }]}}{{dt}} = \frac{{{{[N{a^ + }]}_{bath}} - [N{a^ + }]}}{{{\tau _{N{a^ + }}}}} - 3{I_{p}} + {\delta _{N{a^ + }}}{V_{{K^ + }}}.
\end{eqnarray}

We take the current through the sodium-potassium pump in the form~\cite{Cressman:2009}, including the influence of ATP in the functional dependence:
\begin{eqnarray}
I_{p} = \frac{{\rho {\rm{([ATP])}}}}{{\left( {1 + \exp \left( {3.5 - [{K^ + }]} \right)} \right)\left( {1 + \exp \left( {\frac{{25 - [N{a^{2 + }}]}}{3}} \right)} \right)}}.
\end{eqnarray}
In this case, the excitability of the neuron depends on the potassium concentration:
${V_{{K^ + }}} = \overline \nu ({K^ + })\, + {V_0}_{_{{K^ + }}}$, where $\bar v\left( {{{[K]}_o}} \right)$ defined in~\cite{Cressman:2009}.

Metabolic reserves are determined by the amount of ATP, which depends on the degree of oxygen saturation of the neuron, the current through the sodium-potassium pump and the level of synaptic activity itself:



\begin{eqnarray}
\frac{{d[ATP]}}{{dt}} = \frac{{{{[ATP]}_0} - [ATP]}}{{{\tau _{ATP}}}} +
\end{eqnarray}
\begin{eqnarray*}
+{k_{ATP}} \cdot M{M_s}([{O_2}],{k_{{s_{ATP - O2}}}},{m_{{s_{ATP - O2}}}}) - \\
 - {\gamma _{ATP}}{I_{pump}} - [ATP] \cdot M{M_s}({V_{{K^ + }}},{k_{{s_V}}},{m_{{s_V}}}).
\end{eqnarray*}
In addition to AA metabolites, the vessel radius, taking into account synaptic activity, is also governed by the level of potassium:
\begin{eqnarray}
\frac{{dr}}{{dt}} = \frac{{{r_{rest}} - r}}{{{\tau _r}}} +
\end{eqnarray}
\begin{eqnarray*}
+{k_{EETs}}[EETs] - {k_{HETE}}[HETE] + {k_{{K^ + }}}[{K^ + }].
\end{eqnarray*}
We also include the effects of glutamate and sodium-potassium pump current on cerebral oxygen metabolic rate:


\begin{eqnarray}
{k_{CMR{O_{_2}}}} = {k_{CMRO{2_0}}} + 
\end{eqnarray}
\begin{eqnarray*}
+{k_{{I_{pump}}}} \cdot M{M_s}({I_{pump}},{k_{{s_{Ipump}}}},{m_{{s_{Ipump}}}}){\rm{ + }}\\
+{k_{Glu}} \cdot M{M_s}([Glu],{k_{{s_{Glu}}}},{m_{{s_{Glu}}}}).
\end{eqnarray*}

To describe IP$_3$-dependent calcium dynamics in an astrocyte, we use the model approach from~\cite{Ullah:2006} with modifications described in~\cite{Verisokin:2021}, where calcium in the endoplasmic reticulum, dynamics of IP$_3$, glutamate and all the fluxes included in the equations for cytoplasmic calcium are determined. 


\begin{eqnarray}
\frac{{d{{[C{a^{2 + }}]}_{cyt}}}}{{dt}} = (1 - SVR) \cdot \left( {{J_{channel}} + {J_{leak}} - {J_{pump}}} \right) +
\end{eqnarray}
\begin{eqnarray*}
+ ({J_{in}} - {J_{out}}) + {f_{C{a^{2 + }}}}
\end{eqnarray*}
{Where $J_{channel}$ is $Ca^{2+}$ release depending on the $Ca^{2+}$ and $IP_3$
concentrations, $J_{leak}$ is the $Ca^{2+}$  leaking out from the ER, $J_{pump}$ is the contribution of ER membrane $Ca^{2+}$ pump. $J_{in}$ describes the sum of background constant $Ca^{2+}$ influx and agonist-dependent IP3-stimulated $Ca^{2+}$ influx across plasma membrane from the extracellular space and $J_{out}$ is an extrusion current.}
\begin{eqnarray}
\frac{{d{{[C{a^{2 + }}]}_{ER}}}}{{dt}} = (1 - SVR) \cdot {\upsilon _{ER}},
\end{eqnarray}
\begin{eqnarray}
\frac{{d[I{P_3}]}}{{dt}} = ({J_\delta } + {J_{Glu}}) - \frac{{[I{P_3}] - {k_{I{P_3}_\infty }}}}{{{\tau _r}}},
\end{eqnarray}
{Where $J_{Glu}$ describes the direct effect of extracellular glutamate on additional calcium influx, $J_{\delta }$ is the calcium-induced production of $IP_3$. }
\begin{eqnarray}
\frac{{dh}}{{dt}} = \frac{{{h_\infty } - h}}{{{\tau _h}}},
\end{eqnarray}
\begin{eqnarray}
\frac{{d[Glu]}}{{dt}} = \frac{{{{[Glu]}_{amb}} - [Glu]}}{{{\tau _{Glu}}}} + {k_{{K^ + }}}{V_{{K^ + }}}.
\end{eqnarray}
The production of metabolites is determined by the concentrations of AA and available oxygen:
\begin{eqnarray}
\frac{{d[AA]}}{{dt}} = \frac{{{{[AA]}_0} - [AA]}}{{{\tau _{AA}}}} + {k_{AA}} \cdot MM([{\rm{C}}{{\rm{a}}^{2 + }}]_c^2,\,\,{k_{A}})-\\
 \notag - {k_{E}} \cdot MM(AA,{k_{AE}}) \cdot MM(PO_2,\,\,k_{P{O_2}}) - {\rm{k}}_{{\rm{AA}}{\rm{, b}}}{\rm{[AA]}}{\rm{.}}
\end{eqnarray}
Here and below, the function $MM$ defines a sigmoid-type dependence. Due to the identical nature of the effects of $EETs$ and $PGE_2$ on a blood vessel, we limit ourselves to considering $EETs$, the equation for which has the form
\begin{eqnarray}
\frac{{d[EETs]}}{{dt}}=\frac{{{{[EETs]}_0} - [EETs]}}{{{\tau _{EETs}}}} + ~~~~~~~~~~     \\
 \notag +{k_{E}} \cdot MM(AA,{k_{AE}}) \cdot MM(PO_2,\,\,k_{P{O_2}}) + {f_{EETs}}
\end{eqnarray}

The level of AA in the smooth muscle of a blood vessel is determined by its diffusion from the astrocyte through the intercellular space, as well as the local level of oxygen concentration and is determined by the equation
\begin{align}
\frac{{d{{[AA]}_{ext}}}}{{dt}}{\rm{ =  }}{k_{AA,\,b}}[AA] - {k_{HETE}} \cdot MM(PO_2^2,\,\,k_{P{O_2}, }).
\end{align}

The synthesis of $[20-HETE]$ depends on the concentration of AA in the smooth muscle of the blood vessel and is determined as follows:
\begin{eqnarray}
\frac{{d[HETE]}}{{dt}}{\rm{ =  }}\frac{{{{[HETE]}_0} - [HETE]}}{{{\tau _{HETE}}}} +\\ 
 \notag {k_{HETE}} \cdot MM({[AA]_{ext}},\,\,{k_{AA,\,HETE}}) + {f_{HETE}},
\end{eqnarray}

\section{Numerical solution}
Let us consider the influence of the spatial characteristics of NGVU on the propagation of the CSD {(cortical spreading depression)} wave, taking into account the effect of AA metabolites. To do this, we carry out the solution in a two-dimensional domain, considering that each of its points represents an NGVU, and all the points are connected to each other through diffusion connections.

When constructing the concept of NGVU used in this work, we are based on experimental data~\cite{Volterra:2014},~\cite{Khakh:2015}, according to which the cell bodies of excitatory/pyramidal neurons lie separately from astrocytes; the spatial domain of the astrocyte covers dendrites from a large number of neurons and may contain several bodies of inhibitory neurons; several vessels can pass through one astrocyte domain, each of which in this space will be covered by its “leg”. 
According to this results one astrocyte is associated with 4--10 NGVUs. So if we take the average size of astrocytic cells in mice (approximately 40–60 by 40–60 μm) and the number of neuron axons per this area, then one NGVU will correspond to an area of 20x20 µm. 

Numerical solutions were performed in the AGEOM environment~\cite{AGEOM_CUDA_SOFTWARE}. In this case, the solutions of the proposed model occur on a rectangular template, in which each pixel represents a separate NGVU. At each point in the region, the previously proposed point model of NGVU is solved; the connection between individual pixels is organized through diffusion processes.

\begin{figure}[!htbp]
\centerline{\includegraphics[width=6cm]{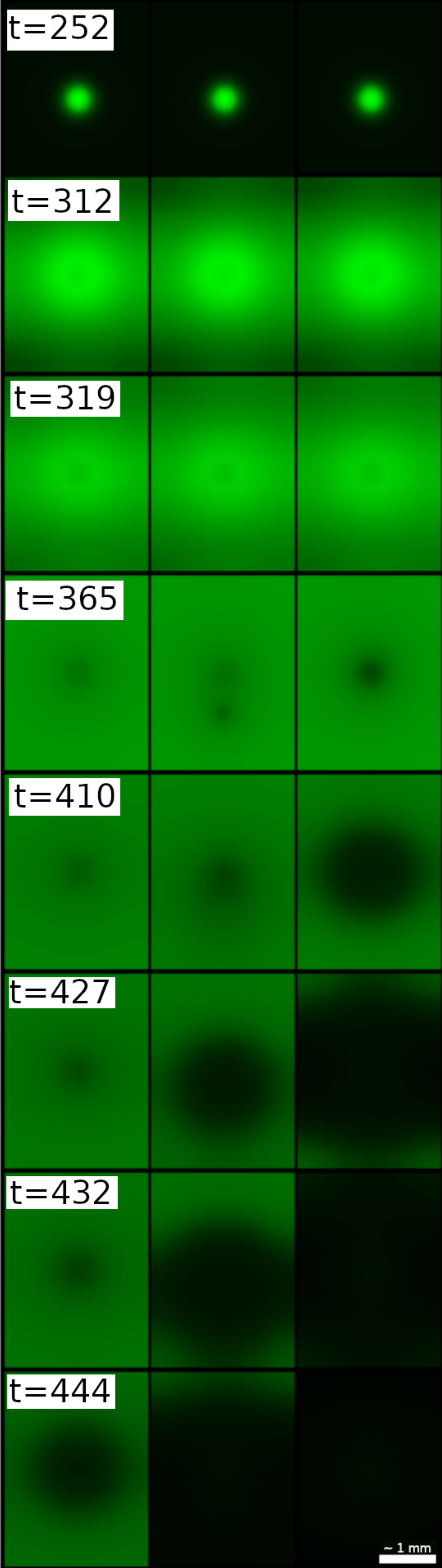}}
\caption{Examples of the occurrence of an CSD wave on a spatial pattern of connected NGVUs. The figures show the dynamics of changes in potassium under the short stimulus occurred in the center of the modeling area. The central column shows a solution with a short change in the level of astrocytic activity near the area of the initial stimulation, in the right column - directly at the site of stimulation}
\label{Solution}
\end{figure}

The solution sequences in Fig.~\ref{Solution} display the occurrence of CSD waves arising in the center of the solution region in connection with external potassium stimulation over a period of time from 250 to 300 seconds. Due to stimulation, we observe a wave that propagates within neighboring NGVUs in space. Fig.~\ref{Solution} on the left shows the emergence and dynamics of the evolution of such a wave in the absence of any other influences. 

To study the influence of AA metabolites on the dynamics of CSD, we changed the intensity of calcium activity by changing the level of ${J_{Glu}}$ over a time period from 380 to 400 s in the areas in which potassium stimulation were applied, as well as at some distance from it. In both cases, a decrease in CSD time is observed. When stimulated in the central solution area (Fig.~\ref{Solution} on the right), exit from the CSD state occurs very quickly, since the increase in ATP quickly brings the NGVU to a normal state. If the stimulation was in the neighboring area, then the recovery delay becomes somewhat longer (Fig.~\ref{Solution} center), but the exit from the CSD state occurs significantly faster than without stimulation.

\section{Conclusions}
The model proposed in the work simulates the key pathways of interaction between the main elements of nervous tissue (astrocytes, neurons, blood vessels, intercellular processes) taking into account the dynamics of AA and its metabolites.

The resulting solution of the spatial model qualitatively and quantitatively corresponds to the experimental data and makes it possible to determine the nature of the influence of AA metabolites on the propagation of CSD waves.
Numerical studies suggest that controlling the dynamics of the trailing edge of the CSD wave in nervous tissue is possible, among other things, by influencing astrocytic activity by activating the synthesis of AA metabolites, which have a direct effect on vascular tone and the release of energy reserves. 

The results obtained can be used both to analyze the ways of interaction between elements of the neurogliovascular unit, and to determine possible therapeutic ways of influencing nervous tissue, the functioning of which is impaired due to neurodegenerative pathologies.


\section*{Acknowledgment}
This study was supported by the Russian Science Foundation, grant 22-74-00146.

\printbibliography

\end{document}